\documentclass{ifacconf}

\usepackage{graphicx}      
\usepackage{natbib}        

\begin{document}
\begin{frontmatter}

\title{VR with Older Adults: \\Participatory Design of a Virtual \\ ATM Training Simulation} 




\author[PJAIT]{Wies\l{}aw Kope\'{c}}
\author[PJAIT]{Marcin Wichrowski}
\author[PJAIT]{Krzysztof Kalinowski}
\author[KOBO]{Anna Jaskulska}
\author[PJAIT]{Kinga Skorupska}
\author[PJAIT]{Daniel Cnotkowski}
\author[PJAIT]{Jakub Tyszka}
\author[PJAIT]{Agata Popieluch}
\author[PJAIT]{Anna Voitenkova}
\author[PJAIT]{Rafa\l{} Mas\l{}yk}
\author[PJAIT]{Piotr Gago}
\author[PJAIT]{Maciej Krzywicki}
\author[SWPS]{Monika Kornacka}
\author[OPI]{Cezary Biele}
\author[OPI]{Pawe{\l} Kobyli{\'n}ski}
\author[OPI]{Jaros{\l}aw Kowalski}
\author[OPI]{Katarzyna Abramczuk}
\author[OPI]{Aldona Zdrodowska}
\author[PAN]{Grzegorz Pochwatko}
\author[PW]{Jakub Mo\.{z}aryn}
\author[PJAIT]{Krzysztof Marasek}

\address[PJAIT]{Polish-Japanese Academy of Information Technology, \\Warsaw, Poland (e-mail: kopec@pja.edu.pl)}
\address[OPI]{National Information Processing Institute, Warsaw, Poland}
\address[KOBO]{Kobo Association, Warsaw, Poland}
\address[PAN]{Institute of Psychology Polish Academy of Sciences, \\Warsaw, Poland}
\address[SWPS]{SWPS University of Social Sciences and Humanities, \\Warsaw, Poland}
\address[PW]{Warsaw University of Technology, Warsaw, Poland}

\begin{abstract}                
In this paper we report on a study conducted with a group of older adults in which they engaged in participatory design workshops to create a VR ATM training simulation. Based on observation, recordings and the developed VR application we present the results of the workshops and offer considerations and recommendations for organizing opportunities for end users, in this case older adults, to directly engage in co-creation of cutting-edge ICT solutions. These include co-designing interfaces and interaction schemes for emerging technologies like VR and AR. We discuss such aspects as user engagement and hardware and software tools suitable for participatory prototyping of VR applications. Finally, we present ideas for further research in the area of VR participatory prototyping with users of various proficiency levels, taking steps towards developing a unified framework for co-design in AR and VR.
\end{abstract}


\end{frontmatter}

\section{Introduction}
Virtual reality, despite its long history, only recently became commercially viable as more manufacturers have taken up the challenge of creating their own solutions and gear. In consequence, multiple sectors started to explore the applications of VR, from the gaming industry, through architecture and even the medical world and industry, for example in machine or car manufacturing. VR, apart from entertainment, is used for interior design, product prototyping or even exposure therapy for treating phobias, as in \cite{gerrit2015development} or coping with emotions, pointed to by \cite{kornacka2016steer}. With the VR solutions becoming more widespread it is important to engage all stakeholders, in particular end users, in their design. In this study we directly engage older adults, whose inclusion in co-design, in light of the recent demographic trends, is very important. 

\begin{figure}
    \includegraphics[width=\linewidth]{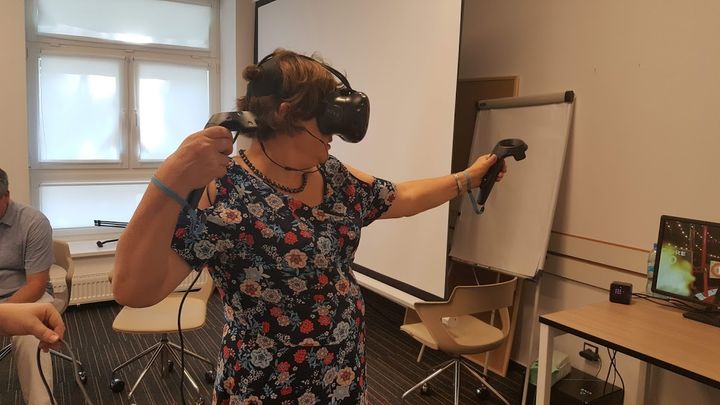}
    \caption{To empower our participants, they were given an opportunity to fully immerse in VR in 360.  A game to play in VR in which they could move around and interact with surrounding environment, among other things, use a bow to shoot at targets.}
    \label{fig:empowerment}
\end{figure}

The data from \cite{population_structure2017} shows that people aged 65+ already comprise about one fifth of the entire EU-28 population (19.2\% in 2016).
Moreover, in most of the western societies the share of older adults is increasing, e.g. in the United States by 2050 it is projected to reach 20\% of the society according to \cite{ortman2014aging}. Thus, in this paper we present our preliminary findings concerning the tools and techniques for involving users of varying ICT-proficiency levels, in our case older adults, in the participatory design process for developing VR solutions. First, we provide an overview of the state of the art. Next, we describe our methods and findings. This section is followed by a discussion listing our preliminary set of best practices for engaging older adults in participatory design of VR solutions. Finally, we summarize our conclusions and propose promising future work areas, in particular the development of a unified comprehensive framework for mixed reality (including VR and AR) co-creation with end users as important actors in the process.

\section{Related work}

Companies, individual developers and researchers have been exploring the potential of VR in different markets, be it online retailing where it is used to supplement the clients' shopping experience as in \cite{bon2018}, or human learning, explored by \cite{su10041102}, and many others. Moreover, as \cite{Busch2014} conclude "virtual environments can be an alternative to real environments for user experience studies, when a high presence is achieved." and as such, they can be used to simulate real experiences and facilitate learning, sometimes referred to as simulation learning, or s-learning. These solutions may be useful for end users, such as older adults, who often are in danger of being stereotyped, but in reality are just as grateful for novel modes of interaction. According to \cite{aula_learning_2004} many older adults value what ICT proficiency can offer them, especially in terms of personal benefits, as shown by \cite{djoub_ict_2013}. However there exist multiple barriers to their use of computer-enabled ICT solutions, noted by \cite{sandhu2013ict}, hence, they are an interesting demographic for promising studies in VR as novel modes of interaction may prove to mitigate some of these existing barriers. To avoid the use of stereotypes in research and design, which were explored by \cite{articleageoldproblem}, it is important to involve the users, as in \cite{sanders2002user} and engage them in participatory development processes, as postulated by  \cite{lindsay2012engaging} and \cite{davidson2013participatory}. Research endeavours ought to provide positive examples of engaging older adults, for example in online crowdsourcing, explored by \cite{skorupska2018smarttv} or offline volunteering, described by \cite{articlefirekeepers}. Such solutions allow older adults' to build on their experience, as explored by \cite{balcerzak2017F1}, while at the same time encouraging them to stay active. For this reason, it is important to get older adults acquainted with technology, and encourage them either with games, as shown by \cite{kopec2017location}, intergenerational interaction explored by \cite{orzeszek2017design} in the context of participatory design and hackathons, as shown by \cite{kopec2017older} allowing them to even build up to being able to join development teams, according to \cite{kopec2017spiral}. Nowadays, this practice should be an industry standard, especially that unexpected insights stemming from user involvement can help avoid business risks according to \cite{pallot2010living}.
This becomes even more important in the context of recent trends such as digital economy and Industry 4.0 related to e.g. EU Digital Single Market policy. Those concepts are connected to numerous emerging aspects that have been already the subject of the study, including by our interdisciplinary team who explored various modes of interaction between humans, machines, systems and interfaces such as VR, as explored by \cite{kobylinski2019vr}, VA, as in \cite{kowalski2019voice} and \cite{biele2019kids}, or personal and industry robots, intelligent or smart homes, cities and workplaces as in \cite{piccarra2016designing}.

Thus, since seniors as a user group may be underrepresented when it comes to ICT solution development, especially cutting-edge, such as VR, it is especially important to directly engage them in participatory design, or as \cite{ladner2015design} states, even Design for User Empowerment, where users could themselves become designers. However, this would require extensive training, which is a resource-intensive process, so another approach is needed, one that would enable them to participate without these limitations.

\begin{figure}
    \includegraphics[width=\linewidth]{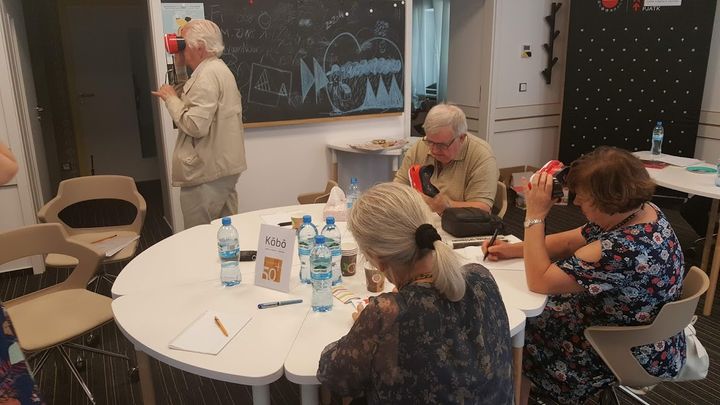}
    \caption{As the first step to develop the VR environment the participants were shown different areas in Google Streetview and could comment on their convenience and security.}
    \label{fig:first}
\end{figure}

\section{Methods}

\subsection{Scenario}

We designed our study to simulate the lean startup environment in which the team works on a real product. This setup allowed us to examine the effectiveness of the applied VR prototyping methods, tools and techniques in the presence of potential users, programmers and researchers in double roles as designers and product owners. The end goal of the development process informed by the workshops, that is the VR cash machine training simulation, was an actual product to be exhibited during Warsaw Senior Days. The scenario of the two VR co-design workshops consisted of multiple steps listed below:

\textbf{Workshop 1 - Introduction and Engagement}
\begin{enumerate}
\item Discussion of the Workshop Goals 
\item Empowerment in VR (Immersion in available VR solutions)
\item VR Development: Environment (Google Street View and discussion to substitute sketching)
\end{enumerate}

\textbf{Workshop 2 - Prototyping and Testing}
\begin{enumerate}
\item UX Prototyping in VR (UI mockup tests in VR)
\item VR Object Prototyping (environment tests in VR)
\item Discussion of the most important conclusions
\end{enumerate}

\subsection{Software and Hardware Setup}
We conducted the initial presentation of VR using \textbf{Disney Movies VR}  featuring 360$^{\circ}$ videos,
to showcase the range of interactions possible in VR we used the \textbf{NVIDIA VR Funhouse} application.
VR UX prototypes were created with the new \textbf{Proto.io VR Prototyping functionality} allowing us to make functional mock-ups of the ATM screen.

\begin{figure}
    \includegraphics[width=\linewidth]{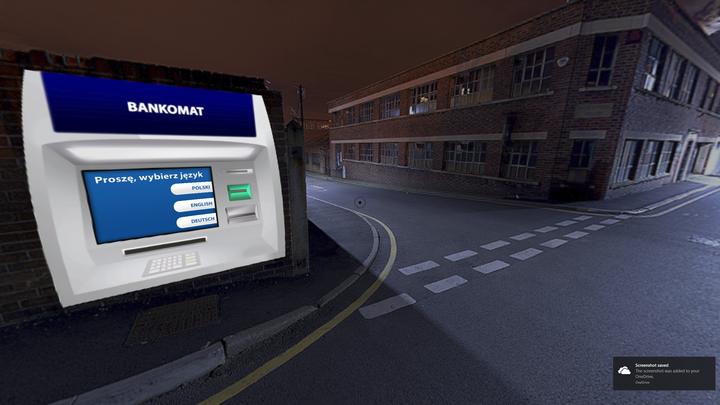}
    \caption{The mock-up in Proto.io's VR prototyping tool allowed us to test the UX in a virtual environment.}
    \vspace{0.45cm}
    \label{fig:proto}
\end{figure}

For scale testing and general testing we used \textbf{Epic Game's Unreal Engine} and due to ease of quick prototyping in it, while the whole project is using \textbf{Unity Game Engine}. 
We chose \textbf{HTC Vive} as the VR platformas it has better future ability to expand to make use of the only eye tracking solution for VR available to us at the moment.
For presenting VR mock-ups from Proto.io as well as Street View in Google Earth VR we used mid-range smartphones which were then inserted into \textbf{Mattel View-Master Viewers}, to allow the users to interact with the screen by clicking a button on the side of the device.

\subsection{Participants}
Our study participants were a group of four retired older adults: two females and two males from our Living Lab (elaborated on in \cite{kopec2017living}). The participants (P1-P4) were selected based on their engagement with the VR technology during our preliminary VR showcase. All of them live in Warsaw, Poland, and they are native Polish speakers. There was a 25 year age span: the youngest participant was 65 years old and the oldest one was 90, with a mean age of 74,55. 

\section{Results}
Here we present the results of our participatory design workshops. Although the participants were engaged in a development process of an ATM VR Simulation, below we list only the insights which are related to the implementation and execution of the workshops, and not the ATM solution itself.

An important insight was the need for controlled immersion in VR: starting with just viewing, without the need to perform complex actions, and with the help of the team. The materials we have chosen for this phase, that is Star Wars and The Jungle Book short VR movies were well received by our participants because of their high quality and immersive experience.

A significant observation relates to our participants' strong need to clearly understand both the purpose and the process of developing the VR solution. Our participants' huge interest was reflected in their preferred use of language and visible in their eagerness to replace unfamiliar words such as "virtual" with familiar terminology, like "simulated" when referring to the solution being developed to enable them to fully participate in the process.

The participants had almost no issues with getting used to VR headset and its controllers. While some of the more unusual concepts were confusing at first  after 1-2 minutes of using the headset everybody had fun and was fairly comfortable with the controls, as shown in Fig. \ref{fig:empowerment}. 
Both the stationary (seated) and room scale (some movement around) VR experiences were treated as novelty by our participants, all of them agreed that room-scale experiences felt more impressive and polished. However, at the same time P3 and P4 concluded that cardboards are better for beginners.

Image and videos in 360, such as the ones in Google Earth VR Street View gave our participants the starting point for the discussion of their preferred environment, which was crucial to allow us to keep later changes to bare minimum. The discussion inspired by Street View in 3D and depicted in Fig. \ref{fig:first} was not limited to the parts of town, or areas in the neighborhood where it is beneficial to place an ATM, but it also included other people who may appear in the virtual space (and some participants dreaded the idea of somebody standing behind them in VR). All of the participants agreed that different environments can be useful for the purposes of the simulation in order to practice different aspects of security and safety when withdrawing money, including any devices that may be attached to ATMs as well as suspicious people lurking in the neighborhood.

After having viewed the VR ATM simulation depicted in Fig. \ref{fig:proto} 
the participants postulated numerous improvements to the UX flow of the UI to improve clarity and speed of interaction, which they deemed as a critical aspect of their ATM experience. One such suggestion was the need for a multi-level interface which hides options, that are not popular with every users to improve the flow for everyday transactions.

\begin{figure}
    \includegraphics[width=\linewidth]{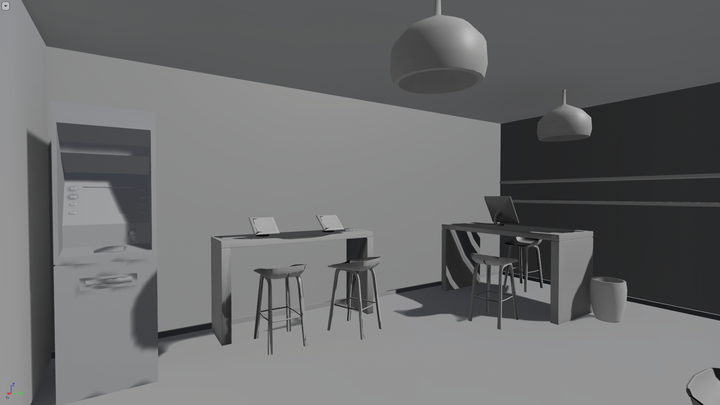}
    \caption{3D model of the ATM was created, to test the dimensions with the participants.}
    \vspace{0.5cm}
    \label{fig:3d}
\end{figure}

When it came to the VR 3D object prototyping, depicted in Fig. \ref{fig:3d},
the participants commented on its placement in space, as well as the scale and placement of the elements such as the card reader or a printer, which they discussed before. Another aspect of the play-space setup was the existence of virtual walls in the VR prototype, as P3 said that a new VR user should start their adventure while sitting down, as they may accidentally want to lean against a non-existent wall. All in all, the need for physical feedback (e.g. with a use of workaround, such as a plank or other props) was often highlighted.

\section{Discussion: Considerations and Recommendations}

\subsection{\textbf{Participants}}
In participatory design high motivation and engagement are crucial, so it is important to assemble the team and user groups who enjoy working together. We found that granting older adults access to technology, which may otherwise be out of their reach is a very good way to guarantee engagement, and usually it is enough to convince them to participate in the development process. Additionally, the whole team and the users alike need to feel that their insights are appreciated and valued, so it is key to schedule the meetings with enough time for digression and extensive questions about the technology or the project itself. 

\subsection{\textbf{Software and Hardware}}

Some participants concluded that cardboards may be a better option as an introduction to VR, as they allow users to stay in control, are easy and intuitive to manage (P1) and can be taken off if such need arises (P3).
Overall, our participants noticed that immersion is fuller when using the stationary VR headset (P2 enjoyed it especially, as they found it to be the most comfortable one). In terms of interactivity, they were divided - as half of them preferred selecting options with a gaze pointer solution, while the others preferred to use controllers. The moment of entry and exit to VR was especially prominent in the standalone VR headset, so the assistance of a facilitator was very welcome. Common suggestion from our participants was to start using VR when sitting down, to avoid the danger of bumping into real things or leaning against virtual objects.

\subsection{\textbf{Methods and Tools}}

All of our software choices met our expectations, and we can especially recommend Google Earth VR Street View with cardboards - some older adults were even curious about how to recreate this experience for themselves later, which also increased their engagement. Proto.io's VR tool was also very effective, however in this case it is important to be aware of the choice of colors and scale, as some elements may not appear as clearly when viewed on smartphone screens. 

The methods and tools we have chosen are listed below:

\begin{itemize}
    \item Brainstorming with mind-mapping
    \item Affinity diagramming
    \item Rating different aspects of presented locations
    \item Cardboard 360 movies and experiences, VR experiences
    \item Commercial VR games and applications, which display different aspects of VR and the range of possible interactions
    \item 360 photos with Google Earth VR Street View in different locations
    \item Proto.io for UX tests on the initial UI
    \item Unreal Engine as a quick means of creating simple VR project
\end{itemize}

While most of the methods we have initially chosen did make their way into the workshops due to time constraints we decided not to add another level of complexity of VR Sketching with Older Adults. Instead, we performed this activity descriptively, agreeing that the development team can create sketches themselves at a later time if needed. In general, the more time is granted to free discussion the more insights can be gathered for UI, UX, models and functionalities, as long as the facilitators aid the users in staying on track by presenting additional elements up for trying out and discussion.

\section{Conclusions and Future Work}

Our research is an important voice in the participatory design landscape, especially that it directly involves end users as co-designers of cutting-edge technologies such as VR. In this study we worked with older adults, who are often underrepresented and affected by stereotypes. We found that engaging them in VR prototyping workshops is a very effective way to directly gain valuable design insights and create, in our case, a useful VR training simulation or in general, any other product. We plan to conduct further research involving end users to facilitate the creation of best VR experiences and to inform the development of a unified and comprehensive co-design framework aimed to deepen the users' participation in the co-design process, especially of mixed reality training simulations. Such framework, having been tested with vulnerable groups, could easily extend to other contexts where stakeholder participation to increase the value proposition is critical, including civil engagement, industry and workplace empowerment. Hence, we are really excited about the prospects of resulting interdisciplinary follow-up discussion on user empowerment and co-creation in VR, especially for s-learning and industry applications.

\begin{ack}
We would like to thank older adults from our Living Lab, those affiliated with Kobo Association who participated in this study and all interdisciplinary experts involved with the HASE Lab group (Human Aspects in Science and Engineering).
\end{ack}

\bibliography{ifacconf}

\end{document}